\documentclass{PoS}
\pdfoutput=1

\usepackage{amssymb}
\usepackage{amsmath}
\usepackage{xcolor}
\usepackage{listings}
\usepackage{booktabs}
\usepackage{siunitx}

\newcommand{\WMG}{\mbox{DD-$\alpha$AMG}}
\newcommand{\CPP}[1][]{C\texttt{++}#1}
\newcommand{\inlinecode}[1]{\lstinline[identifierstyle=\color{black},basicstyle=\ttfamily]{#1}}

\title{pMR: A high-performance communication library\thanks{Work supported
  by the German Research Foundation (DFG) in the framework of SFB/TRR-55.}}

\ShortTitle{pMR: A high-performance communication library}

\author{\speaker{Peter Georg}, Daniel Richtmann, Tilo Wettig\\
  Department of Physics, University of Regensburg, 93040 Regensburg, Germany\\
  E-mail: \email{peter.georg@ur.de}}

\abstract{On  many  parallel  machines,  the time  LQCD  applications  spent  in
  communication  is a  significant contribution  to the  total wall-clock  time,
  especially in  the strong-scaling limit.  We present a  novel high-performance
  communication library that  can be used as a de  facto drop-in replacement for
  MPI  in existing  software. Its  lightweight nature  that avoids  some of  the
  unnecessary overhead introduced by MPI  allows us to improve the communication
  performance   of  applications   without   any   algorithmic  or   complicated
  implementation changes.  As a first real-world  benchmark, we make use  of the
  pMR library in  the coarse-grid solve of the Regensburg  implementation of the
  \WMG\ algorithm. On  realistic lattices, we see an improvement  of a factor 2x
  in pure communication time and total execution time savings of up to 20\%.}

\FullConference{34th annual International Symposium on Lattice Field Theory\\
  24-30 July 2016\\
  University of Southampton, UK}

\begin{document}

\section{Introduction and motivation}
The Regensburg  group (RQCD) has  shown that  their implementation of  the \WMG\
algorithm \cite{Frommer:2013fsa}, targeting the  first-generation Intel Xeon Phi
architecture (a.k.a.\ KNC),  is well optimized in terms of  computation and that
its run time  is now dominated by  off-chip communication \cite{Heybrock:lat15}.
For  off-chip  communication,  our  implementation of  \WMG\  uses  the  Message
Passing  Interface (MPI).  Hence the  performance  depends on  the specific  MPI
implementation  used.  We use  Intel  MPI,  which is  the  de  facto choice  for
applications  running  natively on  the  KNC.  Unfortunately, its  closed-source
character does not allow us  to directly contribute any improvements. Developing
a  new MPI-conformant  library  would involve  major efforts  and  would not  be
sensible anyway since most parts of MPI  are either not required or only used in
parts that are not performance-critical. Contributing to an existing open-source
MPI implementation is  an option, but we  would then have to  strictly adhere to
the MPI standard.  Relaxing this constraint allows  us to design a  new API with
performance benefits  not possible  otherwise. We  therefore decided  to develop
a  novel  high-performance  communication  library well  suited  for  \WMG\  and
stencil-type applications in general, with QPACE~2~\cite{Arts:2015jia} being the
initial target.

\section{Overview of the communication library}
Before  designing   a  new  high-performance  communication   library  some  key
objectives have to be identified. The most obvious is to efficiently utilize all
available resources of the network hardware.  In the case of QPACE~2, which uses
InfiniBand FDR  for off-chip  communication, this  includes using  Remote Direct
Memory Access (RDMA)  capabilities. This is the origin of  the name pico Message
Passing for RDMA (pMR)~\cite{pmr} of  the communication library. Efficient usage
of  InfiniBand hardware  requires  re-using hardware  resources  that have  been
established once  as often as possible.  Hence pMR was designed  with persistent
communication in  mind. RDMA and  persistent communication help to  minimize the
latency imposed by the hardware.  Any additional software-induced latency should
be reduced  to a bare minimum.  To achieve this goal  we adhere to a  few coding
restrictions, e.g., no polymorphism at all to avoid vtable lookups. In addition,
we  do  not  spawn  any  extra  threads within  the  library  to  avoid  context
switches.\footnote{This is in  contrast to many MPI  implementations which spawn
  an arbitrary  number of  threads to  improve performance.  However, experience
  shows  that  these  threads  often interact  with  the  application's  threads
  and  thus  cause  context  switches  and  degrade  performance.  In  some  MPI
  implementations these internal threads are not even pinnable.}

As it is  not feasible to implement  all MPI features and  re-write all existing
software to  use pMR instead of  MPI, it should be  possible to use both  at the
same time and to allow porting to  pMR piece by piece. To facilitate the latter,
the pMR API differs  only slightly from MPI so that  no major code modifications
are  required. Furthermore,  the simplicity  of pMR  allows us  to adapt  to new
networking  hardware and  to support  new  network topologies  with only  modest
efforts.

For the implementation,  we chose \CPP{11} with an optional  C API. An important
objective when  writing the code  was to  minimize dependencies. E.g.,  code for
each  supported  network  provider  is  isolated  to  allow  for  easy  addition
or  removal  without  effects  on  any  other  provider.  Apart  from  improving
maintainability  this allows  for  compile-time optimizations  by only  enabling
required network providers. As a consequence, binaries are cluster specific.

\section{Porting existing software to pMR}
\subsection{General implementation notes}

\begin{figure}[b]
\centering
\noindent\begin{minipage}[t]{.5\textwidth}
\begin{lstlisting}

MPI_Request sendRequest;
MPI_Request recvRequest;

// Setup persistent transfer buffers
MPI_Send_init(sendBuffer, count, MPI_FLOAT, target, sendTag, Comm, sendRequest);
MPI_Recv_init(recvBuffer, count, MPI_FLOAT, target, recvTag, Comm, recvRequest);

for(i = start; i != end; ++i)
{





  // Computation

  MPI_Start(recvRequest);
  MPI_Start(sendRequest);

  // Computation

  MPI_Wait(sendRequest, MPI_STATUS_IGNORE);
  MPI_Wait(recvRequest, MPI_STATUS_IGNORE);

  // Computation
}

MPI_Request_free(recvRequest);
MPI_Request_free(sendRequest);
\end{lstlisting}
\end{minipage}\hfill
    \begin{minipage}[t]{.45\textwidth}
\begin{lstlisting}
// Setup persistent connection
pMR::Connection connection(pMR::Target(
    Comm, target, sendTag, recvTag));

// Setup persistent transfer buffers
pMR::SendWindow<float> sendWindow(
    connection, sendBuffer, count);
pMR::RecvWindow<float> recvWindow(
    connection, recvBuffer, count);

for(i = start; i != end; ++i)
{
  // Computation

  recvWindow.init();
  sendWindow.init();

  // Computation

  sendWindow.post();
  recvWindow.post();

  // Computation

  sendWindow.wait();
  recvWindow.wait();

  // Computation
}
\end{lstlisting}
\end{minipage}

\caption{Halo  exchange using  persistent MPI  point-to-point communication  vs.
  pMR.}
\label{code:persistentmpi-vs-pmr}
\end{figure}

One of  the key objectives  of pMR  is to allow  for easy successive  porting of
existing MPI  C++ (and C) software.  This process depends on  the particular MPI
communication  method.  The only  two  methods  discussed  in this  section  are
persistent and  non-persistent point-to-point  communication. These  two methods
are  probably  the most  common  ones.  The difference  to  pMR  for the  former
is  depicted  in  Fig.~\ref{code:persistentmpi-vs-pmr}.\footnote{Note  that  the
  pseudo-code in Figs.~\ref{code:persistentmpi-vs-pmr} and \ref{code:mpi-vs-pmr}
  is missing proper error handling. For MPI  it is necessary to check the return
  value of  almost all  MPI functions  for success. pMR  is using  C++ exception
  handling, i.e., it throws an exception whenever an error occurs.}
In this  case, two small modifications  are sufficient. First, pMR  requires the
user  to  setup  persistent  connections  to maximize  the  possible  re-use  of
resources. Second, both send and receive routines are split into three functions
instead of two to allow for further overlap of communication and computation. In
both MPI and pMR,  to have persistent send and receive  buffers, we need handles
associated with a particular buffer. These  handles provide all required data to
allow for  re-using resources  for data  transfers working  on the  same buffer.
Hence  a data  transfer is  initiated by  the handle  instead of  the associated
buffer.  pMR uses  different handles  for send  and receive  buffers, i.e.,  the
\inlinecode{SendWindow} and \inlinecode{RecvWindow} classes.

For the case of non-persistent MPI one  more step is required when moving to pMR
(see Fig.~\ref{code:mpi-vs-pmr}): since pMR is  persistent, we need to introduce
handles for  the send and  receive buffers as  explained above. This  step could
have the consequence of  fundamental code changes since we need  to refer to the
handles when performing  data transfers, i.e., these changes are  imposed by the
switch  from  the  non-persistent  to the  persistent  communication  model.  If
desired, the changes can be minimized  at the cost of introducing some buffering
overhead as described below.

\begin{figure}[t]
\centering
\noindent\begin{minipage}[t]{.5\textwidth}
\begin{lstlisting}

MPI_Request sendRequest;
MPI_Request recvRequest;







for(i = start; i != end; ++i)
{





  // Computation

  MPI_Irecv(recvBuffer, count, MPI_FLOAT, target, recvTag, Comm, recvRequest);
  MPI_Isend(sendBuffer, count, MPI_FLOAT, target, sendTag, Comm, sendRequest);

  // Computation

  MPI_Wait(sendRequest, MPI_STATUS_IGNORE);
  MPI_Wait(recvRequest, MPI_STATUS_IGNORE);

  // Computation
}
\end{lstlisting}
\end{minipage}\hfill
\begin{minipage}[t]{.45\textwidth}
\begin{lstlisting}
// Setup persistent connection
pMR::Connection connection(pMR::Target(
    Comm, target, sendTag, recvTag));

// Setup persistent transfer buffers
pMR::SendWindow<float> sendWindow(
    connection, sendBuffer, count);
pMR::RecvWindow<float> recvWindow(
    connection, recvBuffer, count);

for(i = start; i != end; ++i)
{
  // Computation

  recvWindow.init();
  sendWindow.init();

  // Computation

  sendWindow.post();

  recvWindow.post();


  // Computation

  sendWindow.wait();
  recvWindow.wait();

  // Computation
}
\end{lstlisting}
\end{minipage}
\caption{Halo exchange using non-persistent MPI point-to-point vs. pMR.}
\label{code:mpi-vs-pmr}
\end{figure}

\subsection{\boldmath Porting the RQCD \WMG\ implementation to pMR}
We chose  the coarse-grid part  of the \WMG\  algorithm as the  first real-world
benchmark for pMR as it was shown  earlier to be communication bound. We use the
two-level version,  for which the only  operation on the coarse-grid  is a solve
using the FGMRES iterative  solver. The only performance-relevant communications
in FGMRES  are halo  exchanges and global  sums. For now  we only  optimized the
former, which  were previously implemented as  non-persistent MPI point-to-point
communication.  To avoid  major  code changes  due  to persistent  communication
handles, buffering for  send and receive data was introduced,  i.e., the data to
be transferred are  first copied to a registered send  buffer, and received data
are copied  from a registered receive  buffer to the actual  location. This adds
additional overhead, but  for this first benchmark we decided  to pay this price
to keep code changes to a minimum. Incidentally, the coarse-grid solver requires
such buffering for half of its data transfers anyway since it reduces the number
of packages sent over the  network by gather/scatter operations to/from buffers.
To minimize  this overhead  we have first  threaded the  (previously unthreaded)
copy process to/from  send/receive buffers. This is especially  important on the
KNC, where the memory bandwidth scales linearly with the number of active cores.
While this also helps to improve the  performance for MPI, we do not provide any
details on this point. All data shown below are with multi-threaded copy enabled
for both MPI and pMR.

\section{Results}
The   benchmarks   were   run   on    QPACE~2   and   QPACE~B.   Both   machines
use   KNC   coprocessors   and   an  InfiniBand   network;   for   details   see
Table~\ref{table:hardware}  and  Ref.~\cite{Arts:2015jia}.  Software  and  Intel
MPI  details   are  given   in  Table~\ref{table:software},  while   details  of
the  CLS   lattices~\cite{Bruno:2014jqa}  used   in  the   runs  are   shown  in
Table~\ref{table:cls}. For the benchmark runs on QPACE~B, the differences in the
run times  are solely due to  the differences between  MPI and pMR. For  QPACE 2
there is  an additional topology  effect: QPACE~2  uses a novel  topology called
Flexible Block Torus, which Intel MPI is not aware of.

\begin{figure}[ht]
\centering
\includegraphics{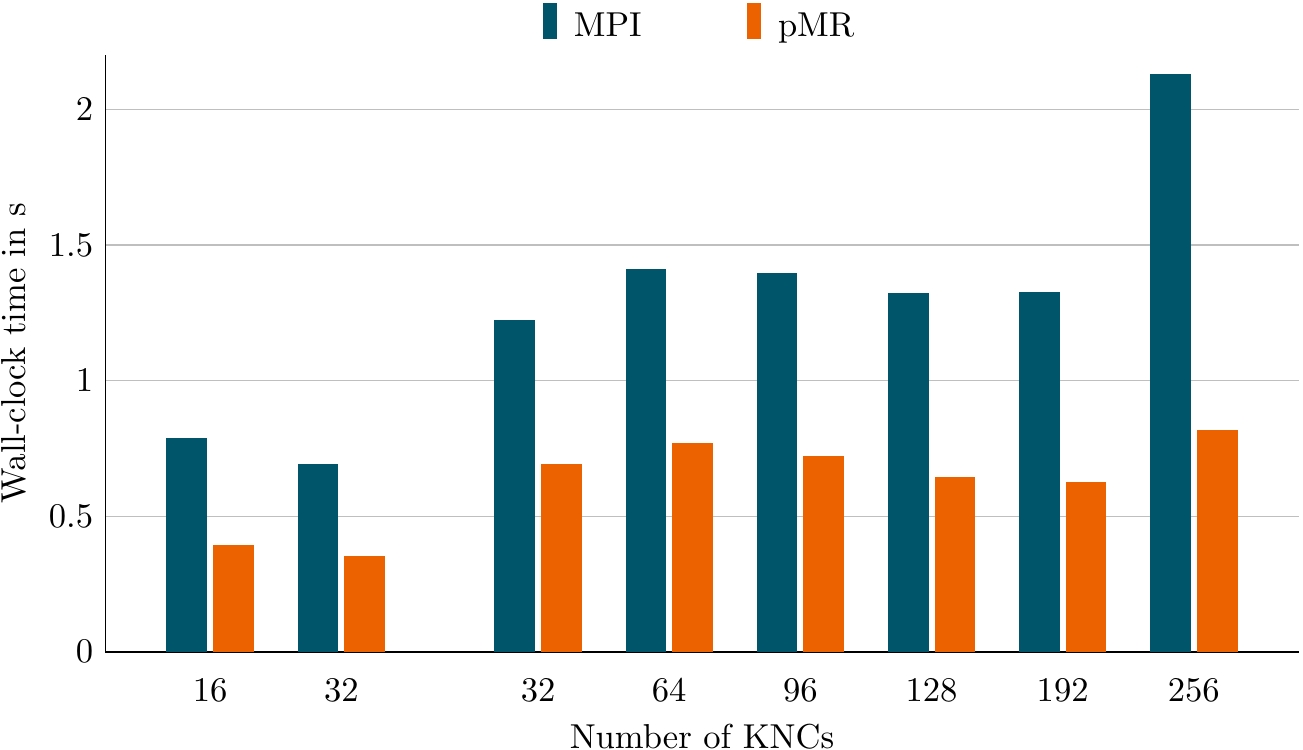}
\caption{Time spent in halo exchanges in the coarse-grid part of a single solve.
  The two left-most pairs of bars are results from QPACE~B, while the others are
  from QPACE~2, with fixed problem size in both cases.}
\label{plot:exchange}
\end{figure}

The results  in Fig.~\ref{plot:exchange} for  the wall-clock time spent  in halo
exchanges (including  copying of data  to/from buffers) shows an  improvement of
about  a factor  2x  for  pMR compared  to  MPI for  both  clusters.  This is  a
significant improvement. The increase for 32 to 64 and 192 to 256 KNCs is due to
the mapping of the lattice  to nodes, see Table~\ref{table:lattice} for details.
In every distributed space-time dimension  (i.e., a dimension involving multiple
nodes) two messages need to be sent  for each halo exchange, i.e., the number of
packages  required per  halo  exchange is  equal to  the  number of  distributed
dimensions times  two. For  a fixed  number of  distributed dimensions  the time
spent in halo  exchanges decreases with the  number of KNCs as  the message size
decreases.

\begin{figure}[ht]
\centering
\includegraphics{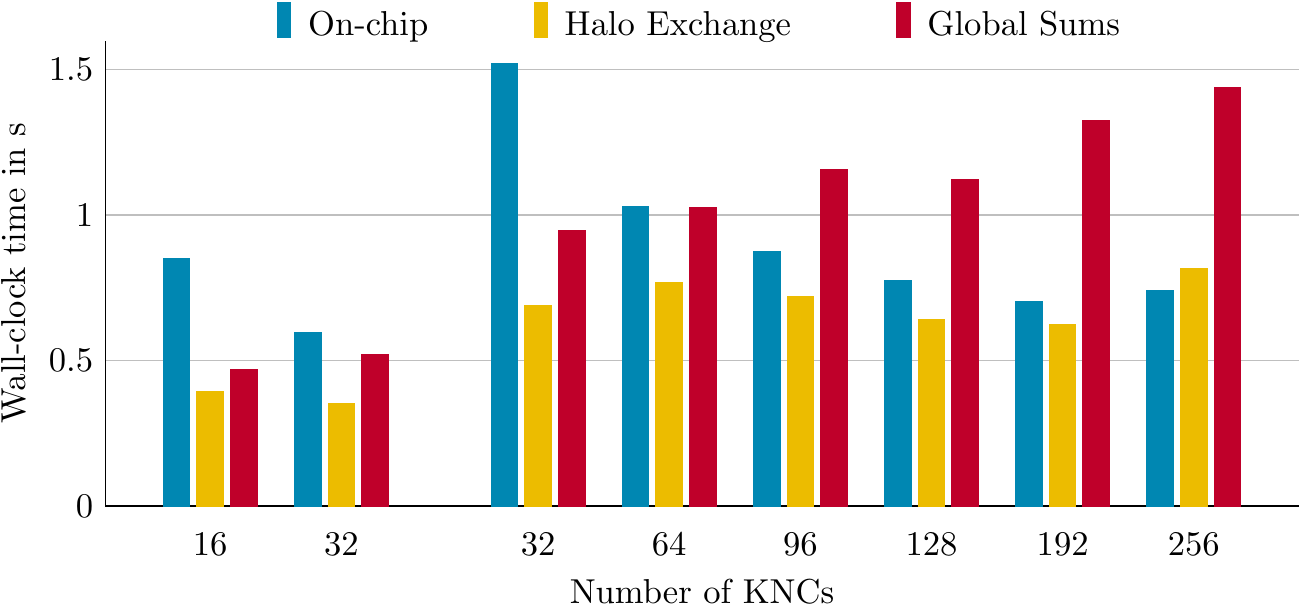}
\caption{Contributions to  run time spent  on the  coarse-grid part of  a single
  solve  using pMR  for halo  exchanges. The  two left-most  groups of  bars are
  results from QPACE~B, the others are  from QPACE~2, with fixed problem size in
  both cases.}
\label{plot:scale}
\end{figure}

Figure~\ref{plot:scale}  shows  the  contribution  of  the  time  spent  on-chip
(computation  and  on-chip synchronization),  halo  exchanges,  and global  sums
for  the coarse-grid  solve,  using pMR  for  halo exchanges  only.  Due to  the
improvements  obtained from  utilizing pMR  for halo  exchanges, the  time spent
on-chip is now almost  equal to the sum of halo exchanges and  global sums for a
low number of  KNCs. Scaling to a  higher number of KNCs at  fixed problem size,
the  global sums  become the  dominant  contribution. Hence,  to achieve  better
strong scaling of the  coarse grid part it is crucial to  improve (or reduce the
number of) global sums. In other words, they are the next target for pMR.

In the case of 256 KNCs there are two interesting observations. First, we see an
increase in the on-chip contribution for this particular fixed problem size. The
reason  is that  the local  problem size  per KNC  is too  small to  efficiently
utilize  the available  resources and  the on-chip  synchronization overhead  is
critical. Second, when running  with MPI only, the time spent  in global sums is
about twice larger  than when running with  pMR for halo exchanges  (but not for
global sums). We do not have an explanation for the second observation.

\section{Conclusions and outlook}
By replacing MPI with  pMR in the halo exchanges of  the \WMG\ coarse-grid solve
we were able to significantly reduce the share of total wall-clock time for this
previously dominant  contribution. We  have therefore  been able  to demonstrate
that communication-bound  stencil-type applications  can be notably  improved by
utilizing our  novel high-performance  communication library without  major code
changes, let alone algorithmic changes.

For  the \WMG\  coarse-grid solve  the global  sums are  now the  dominant part.
Therefore we plan to extend the preliminary global reductions support of pMR and
implement it in  the RQCD \WMG\ implementation. This is  especially important to
improve the coarse-grid  scaling behavior. However, currently  the main priority
is to  add support  for our  new cluster QPACE~3  to pMR.  QPACE~3 is  not using
InfiniBand for inter-node communication, but  Intel Omni-Path, a technology that
is very different to InfiniBand in several aspects.

\appendix
\section{Benchmark details}
\begin{table}[h]
\centering
\begin{tabular*}{\textwidth}{@{\extracolsep{\fill}}*5c@{}}\toprule
          & Host Xeon   & Xeon Phi & InfiniBand     & Topology\\\midrule
  QPACE~2 & E3-1230L v3 & 4x 7120X & Connect-IB FDR & Flexible Block Torus\\
  QPACE~B & E5-2603 v2  & 2x 31S1P & ConnectX-2 QDR & Single switch\\\bottomrule
\end{tabular*}
\caption{Hardware overview.}
\label{table:hardware}
\end{table}

\begin{table}[ht]
\centering
\begin{tabular*}{\textwidth}{@{\extracolsep{\fill}}*6c@{}}\toprule
  OS       & \hspace*{-3.64pt}Intel MPSS & OFED   & Intel MPI    & MPI provider           & Allreduce\\\midrule
  CentOS 7 & 3.5.1                       & 3.12-1 & 5.1 Update 3 & DAPL (CCL-direct only) & Recursive doubling\\\bottomrule 
\end{tabular*}
\caption{Software overview.}
\label{table:software}
\end{table}

\begin{table}[ht]
\centering
\begin{tabular*}{\textwidth}{@{\extracolsep{\fill}}*7c@{}}\toprule
          & id   & $\beta$ & $N_s$ & $N_t$ & $m_\pi$                        & $a$\\\midrule
  QPACE~2 & C101 & $3.4$   & $48$  & $96$  & $\SI{220}{\mega\electronvolt}$ & $\SI{0.086}{\femto\meter}$\\
  QPACE~B & H102 & $3.4$   & $32$  & $96$  & $\SI{350}{\mega\electronvolt}$ & $\SI{0.086}{\femto\meter}$\\\bottomrule
\end{tabular*}
\caption{CLS lattice configurations used for benchmarks \cite{Bruno:2014jqa}.}
\label{table:cls}
\end{table}

\begin{table}[ht]
\centering
\begin{tabular*}{\textwidth}{@{\extracolsep{\fill}}*7c@{}}\toprule
$16$ & $32$ & $64$ & $96$ & $128$ & $192$ & $256$\\\midrule
$1\!\times\!1\!\times\!2\!\times\!8$ & $1\!\times\!1\!\times\!4\!\times\!8$ & $1\!\times\!4\!\times\!4\!\times\!4$ & $1\!\times\!4\!\times\!3\!\times\!8$ & $1\!\times\!4\!\times\!4\!\times\!8$ & $1\!\times\!4\!\times\!4\!\times\!12$ & $2\!\times\!4\!\times\!4\!\times\!8$\\\bottomrule
\end{tabular*}
\caption{Mapping of the lattice (bottom) to number of KNCs (top). The RQCD \WMG\
  implementation imposes some  restrictions on the possible  mappings. We picked
  the mappings that minimize the total execution time for MPI.}
\label{table:lattice}
\end{table}

\bibliographystyle{jbJHEP}
\bibliography{references}

\end{document}